\documentclass[twocolumn,preprintnumbers,amsmath,amssymb,pra]{revtex4}

\usepackage{graphicx}
\usepackage{dcolumn}
\usepackage{bm}
\usepackage{float}

\usepackage{amsmath}
\usepackage[T2A]{fontenc}
\usepackage[utf8]{inputenc}

\begin{document}

\title{
Quantum phase measurement of two-qubit states in an open waveguide}

\author{Ya. S. Greenberg}\email{yakovgreenberg@yahoo.com}
\affiliation{
 Novosibirsk State Technical University, Novosibirsk, Russia
}

\author{A. A. Shtygashev}
\affiliation{
 Novosibirsk State Technical University, Novosibirsk, Russia
}

\date{\today}

\begin{abstract}
We present a new method for quantum state tomography within a
single-excitation subspace of two-qubit states in an open
waveguide. The system under investigation consists of three qubits
in an open waveguide, separated by a distance comparable to the
wavelength of the electromagnetic field. We show that the
modulation of the frequency of the central ancillary qubit allows
us to obtain unambiguous information about the initial phase
difference $\varphi_1-\varphi_3$ of the edge qubits via the
measurement of the evolution of their probability amplitudes.

\begin{description}
\item[Keywords]
open quantum systems, two-qubit states, quantum state tomography, waveguides
\end{description}
\end{abstract}

\maketitle

\section{\label{Intro}Introduction}

Extracting information about the quantum state is an essential
task in the benchmarks of quantum devices or quantum information
algorithms. This is referred to as  quantum state tomography
(QST). As in the classical tomography, when we reconstruct a three
dimensional image of the object by the use of its various
projections on a two-dimensional plane, quantum state tomography
reconstructs the state by the use of sequences of quantum gates
and projective measurements \cite{Toninelli2019}. A consequence of
projective measurements is that the state is destroyed, therefore
these sequences should be implemented onto a set of identical
quantum systems or onto the same system prepared repeatedly in the
same state \cite{Schmied2016}. In circuit Quantum Electrodynamics
(cQED) one can perform directly measurements in the energy basis
of qubits, or equivalently, measurement of the z-projection on the
Bloch sphere.  These measurements are typically dispersive-shift
based, where the resonance frequency of the readout resonator is
qubit-state dependent \cite{Blais2004}. To obtain the two
remaining projections, one implements $X$ and $Y$ gates prior to
the measurement \cite{Steffen2006}. To reconstruct the state of a
single qubit at least three different gates are needed, and the
density matrix has three independent elements that can be
reconstructed using the measurement results. For two qubits the
problem is already considerably more resource-demanding, as the
number of gates increases  to 9 for  a two qubit state, and the
full density matrix has 15 independent elements that have to be
determined \cite{Wallraff2006}.

Open quantum systems without additional resonators  are  of the special interest
both experimentally \cite{Brehm2021, Forn-Diaz2017, Koshino2013, Mirhosseini2019} and theoretically \cite{Kornovan2015, Fang2014, Fang2015, Issah2021, Kockum2018, Albrecht2019,Greenberg2017,Sultanov2018}. In these systems interference effects appear when the distance between qubits is comparable to a characteristic wavelength. The interference is caused by the effective interaction between the qubits via virtual photons. There are several theoretical works devoted to mentioned interference effects \cite{Greenberg2015,vanLoo2013}, synchronization and superradiance \cite{Cattaneo2021},
as well as  experimental realizations of long-distance interacting superconducting qubits \cite{Wen2019,Zhong2019}.

Here we investigate an open quantum system consisting of an open waveguide, two main qubits and one ancillary central qubit, and we restrict the Hilbert space to a  single-excitation subspace. By employing frequency modulation of the ancillary qubit \cite{Silveri2017} we obtain a one-to-one mapping between the phase of the two qubit off-diagonal density matrix element in the single-excitation subspace  and the measurement result in the energy basis. Thus, the quantum state could be reconstructed by two measurements:
the $\sigma_z$-components of the two qubits without modulation, to get the absolute values of the amplitude probabilities;
and the $\sigma_z$-components of two qubits with modulation, to get the phases of the amplitude probabilities.

In contrast to a common practice where for tomography
reconstruction the gate pulses are applied to the measured qubits,
in our method the measurement pulse is applied to the ancillary
qubit. Until the projective measurements two qubits do not undergo
any external influence.

The paper is structured as follows.

In Section \ref{problem} we obtain the time-dependent differential equations for
the probability amplitudes $\beta_{1,2,3}(t)$ of the three qubits, which account
for the modulation of the frequency  of the central qubit.

The main results of the paper are described in Section \ref{tomo}. In Subsection
\ref{tomoa} we consider the free evolution of three-qubit system. We show that
the free evolution probabilities $|\beta_1(t)|^2$ and $|\beta_3(t)|^2$ depend
on the phase difference $\varphi_1-\varphi_3$. However, the population difference
 $|\beta_1(t)|^2-|\beta_3(t)|^2$  is  phase independent. It is shown that from
 free evolution measurements we can find both the initial values of probability
 amplitudes $\beta_{1,3}(0)$ and the quantity $\cos{(\varphi_1-\varphi_3)}$.
In Subsection \ref{tomob} we consider the solution of the equations obtained in
Section \ref{problem}  under frequency modulation, $f(t)\neq 0$ with the initial
conditions $\left. {\beta _1 (t)} \right|_{t = 0}  = \beta _1 (0);\;\left. {\beta _2 (t)} \right|_{t = 0}  = 0;\;\left. {\beta _3 (t)} \right|_{t = 0}  = \beta _3 (0)$.
From the results obtained in this section, we may conclude that modulating the
frequency of the second qubit allows us to obtain
unambiguous information about the initial phase difference $\varphi_1-\varphi_3$
via the measurement of the evolution of the probability amplitudes
$|\beta_1(t)|^2$, $|\beta_3(t)|^2$.

\section{Formulation of the problem}\label{problem}

We consider a linear chain of three equally spaced qubits which
are coupled to the photon field in an open waveguide (see Fig. 1).

\begin{figure}
 \includegraphics[width=6 cm]{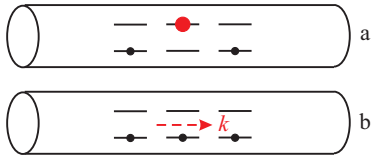}\\
 \caption{Schematic illustration of the single-excitation subspace for
a three-qubit chain in an open waveguide. (a) One qubit is excited, while the other two
qubits are in the ground state. (b) All three qubits are in the ground state and
 a single photon propagates in the waveguide.}\label{FIG:Fig1}
\end{figure}

The distance between neighboring qubits is equal to $d$. The Hilbert
space of each qubit is spanned by the excited state vector $|e\rangle$ and
the ground state  vector $|g\rangle$. The Hamiltonian that accounts for
the interaction between qubits and the electromagnetic field is as
follows (we use $\hbar=1$ throughout the paper):
\begin{equation}\label{eq:math:1}
  H = H_0  + \sum\limits_k {\omega _k a_k^ +  a_k }+H_{int}
\end{equation}
where $H_0$ is the Hamiltonian of the bare qubits and $H_{\operatorname{int} }$ is the interaction Hamiltonian between the qubits and the photons in the waveguide
\begin{equation}\label{eq:math:a2}
{H_0} = \frac{1}{2}\Omega \sum\limits_{n = 1}^3 {\left( {1 + \sigma _z^{\left( n \right)}} \right)}  + \frac{1}{2}f\left( t \right)\left( {1 + \sigma _z^{\left( 2 \right)}} \right),
\end{equation}
\begin{equation}\label{eq:math:a1}
 H_{\operatorname{int} }
 = \sum\limits_{n = 1}^3 {} \sum\limits_k {} g_k^{(n)} e^{ - ikx_n }
 \sigma _ - ^{(n)} a_k^ +   + h.c..
\end{equation}

In Eq. (\ref{eq:math:a2}) the two edge qubits have equal frequencies,
$\Omega$, while the frequency of a central qubit, $\Omega_C(t)$
may be time-dependent: $\Omega_C(t) =f(t)+\Omega$, i.e. detuned by $f(t)$ from the edge qubits. The quantity
$g_k^{(n)}$  in Eq. \eqref{eq:math:a1} denotes the coupling between $n$-th
qubit and the photon field, while $x_n$ is the position of $n$-th qubit.

Below we consider a single-excitation subspace with either a
single photon in the waveguide and all qubits in the ground
state, or with no photons in a waveguide and only one qubit
in the chain being excited. The Hamiltonian Eq. \eqref{eq:math:a1}
conserves the number of excitations (number of excited qubits +
number of photons). Therefore, at any instant of time the system will remain
within the single-excitation subspace. The wave function of an
arbitrary single-excitation state can then be written in the form:
\begin{equation}\label{eq:math:a4}
\left| \Psi (t)  \right\rangle  = \sum\limits_{n = 1}^3 {\beta _n
(t)e^{ - i\Omega t} } \left| {n,0_k } \right\rangle  +
\sum\limits_k {\gamma _k (t)e^{ - i\omega _k t} } \left| {G,1_k }
\right\rangle ,
\end{equation}
where $\beta_n(t)$  is the amplitude of $n$-th qubit, $\left|
{G,1_k } \right\rangle  = \left| {g_1 ,g_2 ,g_3 } \right\rangle
\otimes \left| {1_k } \right\rangle$, $\left| {1,0_k }
\right\rangle  = \left| {e_1 ,g_2 ,g_3 } \right\rangle  \otimes
\left| {0_k } \right\rangle$, $\left| {2,0_k } \right\rangle  =
\left| {g_1 ,e_2 ,g_3 } \right\rangle  \otimes \left| {0_k }
\right\rangle$, $\left| {3,0_k } \right\rangle  = \left| {g_1 ,g_2
,e_3 } \right\rangle  \otimes \left| {0_k } \right\rangle$, and
$\gamma_k(t)$  is a single-photon probability amplitude which is related to a
spectral density of spontaneous emission.

The equations for the amplitudes $\beta_n(t)$ and $\gamma_k(t)$ in
Eq. \eqref{eq:math:a4} can be found from the time-dependent Schr\"odinger
equation $id|\Psi\rangle/dt=H|\Psi\rangle$. For the probability amplitudes
$\beta_n(t)$ we obtain the following equations (the details of the
derivation are given in Appendix A):
\begin{equation}\label{5}
\begin{gathered}
  \frac{{d\beta _1 }}
{{dt}} =  - \frac{\Gamma }
{2}\left( {\beta _1  + \beta _2 e^{ikd}  + \beta _3 e^{i2kd} } \right) \hfill \\
  \frac{{d\beta _2 }}
{{dt}} =  - if(t)\beta _2 (t) - \frac{\Gamma }
{2}\left( {\beta _1 e^{ikd}  + \beta _2  + \beta _3 e^{ikd} } \right) \hfill \\
  \frac{{d\beta _3 }}
{{dt}} =  - \frac{\Gamma }
{2}\left( {\beta _1 e^{i2kd}  + \beta _2 e^{ikd}  + \beta _3 } \right), \hfill \\
\end{gathered}
\end{equation}
where $k=\Omega/v_g$ and $\Gamma$ is the rate of spontaneous
emission of qubit into the waveguide mode.

The wave function which describing the dynamic evolution of the $\beta_n(t)$'s is the projection of the single- excitation
wavefunction Eq. (\ref{eq:math:a4}) on the vacuum photon state:
\begin{equation}\label{6}
\left| \Psi(t)  \right\rangle _0  = \left\langle {0_k |\Psi(t) }
\right\rangle  = \sum\limits_{n = 1}^3 {\beta _n (t)} \left| n
\right\rangle .
\end{equation}
where $\left| {\left. n \right\rangle } \right. = \left\langle {\left. {{0_k}} \right|} \right.\left. {n{{,0}_k}} \right\rangle$ describes the state with $n$-th qubit excited.

We consider the initial state in the following form:
\begin{equation}\label{7}
\left| {\Psi (0)} \right\rangle _0  = \left| {\beta _1 (0)}
\right|e^{i\varphi _1 } \left| 1 \right\rangle  + \left| {\beta _3
(0)} \right|e^{i\varphi _3 } \left| 3 \right\rangle ,
\end{equation}
therefore the second (central) qubit is initially not excited.

In Eq. (\ref{7}) $|\beta_1(0)|, |\beta_3(0)|$ determine the probability to find the 1-st and 3-rd qubit respectively in an excited state, and $\varphi_1, \varphi_3$ are the phases of the amplitude probabilities of these qubits. By definition, this two-qubit state is described by the density matrix:
\begin{equation}\label{rho}
    \rho (0)  = \left( {\begin{array}{*{20}{c}}
{{|\beta_1(0)| ^2}}&{|\beta_1(0)|| \beta_3(0)| {e^{i\left( {{\varphi _1} - {\varphi _3}} \right)}}}\\
{|\beta_1(0)|| \beta_3(0)| {e^{ - i\left( {{\varphi _1} - {\varphi _3}} \right)}}}&{{|\beta_3(0)| ^2}}
\end{array}} \right)
\end{equation}
The aim of tomography is to obtain all the elements of the density matrix.
Here we suppose that one can measure $|\beta_1(0)|, |\beta_3(0)|$, {\it i.e.} $z$-component for each qubit.  The only left component is the phase
difference $\varphi_1-\varphi_2$ and finding it is the centerpiece of our proposal.

In what follows we show that modulating the frequency of the second qubit \cite{Silveri2017} allows for the extraction of the information about the initial values $|\beta_1(0)|$,  $|\beta_3(0)|$, and about the phase difference
$\varphi_1-\varphi_3$ via the measurement of the probability
amplitudes $|\beta_1(t)|^2$, $|\beta_3(t)|^2$. In a typical circuit QED setup, the frequency modulation is realized by varying the current through a line used to produce a bias magnetic field.

\section{Tomography of the two-qubit state}\label{tomo}
\subsection{Free evolution of the three-qubit system}\label{tomoa}
We consider first the solution of Eqs. (\ref{5}) in the absence of a modulation signal, $f(t)=0$ with the initial conditions $\left. {\beta _1 (t)} \right|_{t = 0}  = \beta _1 (0);\;\left. {\beta _2 (t)} \right|_{t = 0}  = 0;\;\left. {\beta _3 (t)} \right|_{t = 0}  = \beta _3 (0)$. For this case, we  obtain for $kd=2\pi$ the following solution:
\begin{eqnarray}
\beta _1 (t)=\frac{1}
{3}\left( {\beta _1 (0) + \beta _3 (0)} \right)e^{ - \frac{{3\Gamma }}
{2}t}  + \frac{2}
{3}\beta _1 (0) - \frac{1}
{3}\beta _3 (0),\label{fr1}  \\
\beta _2 (t) = \frac{1}
{3}\left( {\beta _1 (0) + \beta _3 (0)} \right)e^{ - \frac{{3\Gamma }}
{2}t}  - \frac{1}
{3} {\beta _1 (0)  - \frac{1}
{3}\beta _3 (0)}, \label{fr2}\\
  \beta _3 (t) = \frac{1}
{3}\left( {\beta _1 (0) + \beta _3 (0)} \right)e^{ - \frac{{3\Gamma }}
{2}t}  - \frac{1}
{3}\beta _1 (0) + \frac{2}
{3}\beta _3 (0).\label{fr3}
\end{eqnarray}

Neglecting the first decaying terms in right hand side of equations (\ref{fr1})-(\ref{fr3}) for the time $t> t_0$ where $\Gamma t_0\gg 1$, we obtain
\begin{equation}\label{cos}
   \begin{gathered}
  \left| {\beta _1 (t)} \right|^2  = \frac{1}
{9}\left( {4\left| {\beta _1 (0)} \right|^2  + \left| {\beta _3 (0)} \right|^2 } \right) \hfill \\
   - \frac{4}
{9}\left| {\beta _1 (0)} \right|\left| {\beta _3 (0)} \right|\cos \left( {\varphi _1  - \varphi _3 } \right) \hfill \\
  \left| {\beta _2 (t)} \right|^2  = \frac{1}
{9}\left( {\left| {\beta _3 (0)} \right|^2  + \left| {\beta _1 (0)} \right|^2 } \right) \hfill \\
   + \frac{2}
{9}\left| {\beta _1 (0)} \right|\left| {\beta _3 (0)} \right|\cos \left( {\varphi _1  - \varphi _3 } \right) \hfill \\
  \left| {\beta _3 (t)} \right|^2  = \frac{1}
{9}\left( {4\left| {\beta _3 (0)} \right|^2  + \left| {\beta _1 (0)} \right|^2 } \right) \hfill \\
   - \frac{4}
{9}\left| {\beta _1 (0)} \right|\left| {\beta _3 (0)} \right|\cos \left( {\varphi _1  - \varphi _3 } \right) \hfill \\
\end{gathered}
\end{equation}

\begin{equation}\label{fr4}
 \begin{gathered}
  \left| {\beta _1 (t)} \right|^2  - \left| {\beta _3 (t)} \right|^2  = \frac{1}
{3}\left( {\left| {\beta _1 (0)} \right|^2  - \left| {\beta _3 (0)} \right|^2 } \right) \hfill \\
\end{gathered}
\end{equation}
It follows from Eq. (\ref{fr4}) that if initially $|\beta_1(0)|=|\beta_3(0)|=1/\sqrt{2}$, then at any time $|\beta_1(t)|=|\beta_3(t)|$.

While the evolution of $|\beta_1(t)|^2$ and $|\beta_3(t)|^2$ each depend on the phase difference $\varphi_1-\varphi_3$, their difference is phase independent as seen from Eq. (\ref{fr4}).

Therefore, from the normalization condition $|\beta_1(0)|^2+|\beta_3(0)|^2=1$, we obtain from Eq. (\ref{fr4})
$|\beta_1(0)|^2=\frac{1}{2}(1+3d(t_{0}))$, $|\beta_3(0)|^2=\frac{1}{2}(1-3d(t_{0}))$,  where the measured quantity $d(t_{0})=\left| {\beta _1 (t_{0})} \right|^2  - \left| {\beta _3 (t_{0})} \right|^2$. Then, from any of the Eqs. (\ref{cos}) we can obtain $\cos{(\varphi_1-\varphi_3)}$.

However, an unambiguous knowledge of the phase difference  would require some additional information, for example the value of $\sin{(\varphi_1-\varphi_3)}$. In the following subsection we show that this quantity can be obtained by the frequency modulation of initially not excited central qubit.

\subsection {Measurement of the phase difference by frequency modulation}\label{tomob}

Next we consider the solution of Eqs. (\ref{5}) under frequency modulation, $f(t)\neq 0$ with the initial conditions $\left. {\beta _1 (t)} \right|_{t = 0}  = \beta _1 (0);\;\left. {\beta _2 (t)} \right|_{t = 0}  = 0;\;\left. {\beta _3 (t)} \right|_{t = 0}  = \beta _3 (0)$.

Solving Eqs. (\ref{5}) for $kd=2\pi$
yields the following results (the details of the derivation are
given in Appendix B):
\begin{equation}\label{8}
\begin{gathered}
  \left| {\beta _1 (t)} \right|^2  - \left| {\beta _3 (t)} \right|^2  =
  \frac{1}{3}e^{ - \Lambda (t)}
[
\left({\left| {\beta _1 (0)} \right|^2  - \left| {\beta _3 (0)} \right|^2 } \right) \cos u(t)      \\
    +2\left| {\beta _1 (0)} \right| \left| {\beta _3 (0)} \right| \sin (\phi_1-\phi_3) \sin u(t)
]
\end{gathered}
\end{equation}

\begin{equation}\label{8_a}
 \begin{gathered}
  \left| {\beta _1 (t)} \right|^2  + \left| {\beta _3 (t)} \right|^2  = \frac{1}
{{18}}\left( {e^{ - 2\Lambda (t)}  + 9} \right) \hfill \\
   + \frac{1}
{18}\left( {\beta _1^ *  (0)\beta _3 (0) + \beta _1 (0)\beta _3^ *(0)  } \right)\left( {e^{ -2 \Lambda (t)}  - 9} \right) \hfill \\
\end{gathered}
\end{equation}

\begin{equation}\label{8b}
\left| {\beta _2 (t)} \right|^2  = \frac{1}
{9}e^{ - 2\Lambda (t)} \left( {1 + 2\left| {\beta _1 (0)} \right|\left| {\beta _3 (0)} \right|\cos \left( {\varphi _1  - \varphi _3 } \right)} \right)
\end{equation}
where

\begin{equation}\label{9}
u(t) = \frac{2} {3}\int\limits_0^t {f(\tau )d\tau }
\end{equation}

\begin{equation}\label{10}
\Lambda (t) = \frac{4} {{27\Gamma t}}\left( {\int\limits_0^t
{f(\tau )d\tau } } \right)^2=\frac{1}{3\Gamma t}{u^2(t)}
\end{equation}

It worth noting that Eqs. (\ref{8}) and (\ref{8_a}) are found for $\Gamma t \gg |F(t)|$ or equivalently $\Gamma\gg\Delta\Omega$, where $\Delta\Omega$ is the deviation of the frequency of a second qubit from that of the edge qubits.
From the formal point of view, it means that  the quantity $\Lambda(t)\ll 1$ and in (\ref{B22}) we neglect the decaying exponent $ e^{\lambda _2 }  \approx e^{ - \frac{3} {2}\Gamma t} $. Also, from Eqs. (\ref{9}) and (\ref{10}) , one sees that the dynamics is defined only by the area under the time-function $f(\tau )$. When $f(\tau )\neq 0$, periodic oscillations exist, see Eq. (\ref{8}), with a time-dependent decay rate $\Lambda(t)\ll 1$.
As soon as the detuning between the central qubit and the side qubits goes to zero ($f(\tau )=0$) the integral value becomes constant and the oscillatory dynamics stops. In this sense, $f(\tau )$ could be any arbitrary non-breaking function.

In principle, Eqs.  (\ref{8}), (\ref{8_a}) allow us to obtain both the initial probability amplitudes $\beta_1(0), \beta_3(0)$ and the phase difference $\varphi_1-\varphi_3$. For a $\pi$ pulse ($u(t_{\pi})=\pi$) we obtain from (\ref{8})
\begin{equation}\label{8a}
d(t_{\pi})= -\frac{1}{3}\left( {\left| {\beta _1 (0)} \right|^2  - \left| {\beta _3 (0)} \right|^2 } \right)
\end{equation}
where the measured quantity is the population difference $d(t_{\pi})=\left| {\beta _1 (t_{\pi})} \right|^2  - \left| {\beta _3 (t_{\pi})} \right|^2$.
Together with normalizing condition $|\beta_1(0)|^2+|\beta_3(0)|^2=1$ we obtain from Eq. (\ref{8a})
$|\beta_1(0)|^2=\frac{1}{2}(1-3d(t_{\pi}))$, $|\beta_3(0)|^2=\frac{1}{2}(1+3d(t_{\pi}))$.
We then repeat the measurements for the same initial conditions by applying a $\pi/2$ pulse ($u(t_{\pi/2})=\pi/2$). We obtain
\begin{equation}\label{11}
d(t_{\pi/2})
=  \frac{2} {3}|\beta_1(0)||\beta_3(0)|\sin (\varphi _3  - \varphi _1 )
\end{equation}

\begin{figure}
\centering
    \includegraphics[width=0.49\textwidth]{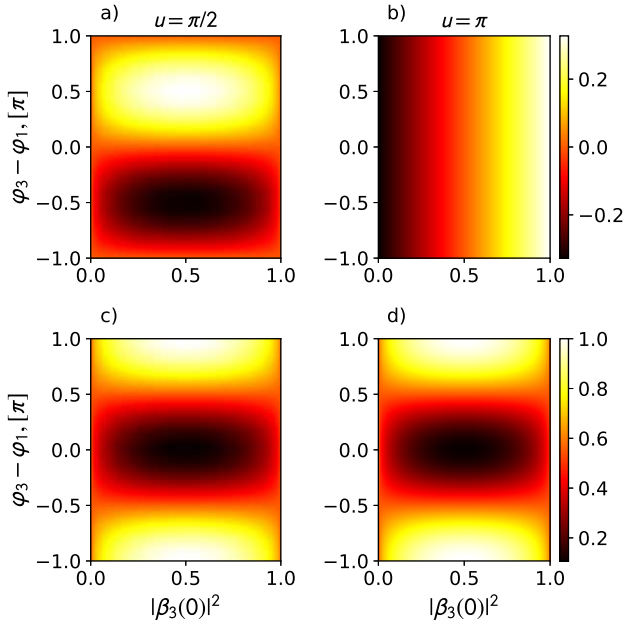}
    \caption[width = 0.49\textwidth]{ Population difference (a,b) and
    sum (c,d) dependence on the initial state parameters  (populations
    and phase difference) after modulation pulses columnwise corresponding
    to $u(t)=\pi /2$ and $u(t)=\pi$. }
    \label{fig:Amp_vs_phase_tomo}
\end{figure}

In Eq. (\ref{11}) the amplitudes $\beta_1(0), \beta_3(0)$ can be obtained either from the free evolution (subsection A) or from  the $\pi$ pulse measurements in Eq.  (\ref{8}). Therefore, the quantity $\sin{(\varphi_1-\varphi_3)}$  is obtained from Eq. (\ref{11}).
In order to obtain the phase difference $\varphi_1-\varphi_3$ unambiguously we may use Eq. (\ref{8_a}) which, under the assumption $\Lambda\ll 1$, can be written as
\begin{equation}\label{8_b}
 S(t_{\pi/2} ) = \frac{5}
{9} - \frac{{8}}
{9}\left| {\beta _1 (0)} \right|\left| {\beta _3 (0)} \right|\cos \left( {\varphi _1  - \varphi _3 } \right)
\end{equation}
where  $S(t_{\pi/2})=\left| {\beta _1 (t_{p})} \right|^2  + \left| {\beta _3 (t_{p})} \right|^2$.

From Eq. (\ref{8}) we see that the measurable value $\left| {\beta _1 (t)} \right|^2  - \left| {\beta _3 (t)} \right|^2$ presents a mix of two types of information. The first term depends only on the initial population difference, while the phase information is contained in the second term. Moreover, $\Lambda$ characterizes the information leak rate from the system to the measurable value. So, at $t=0$ the exponent $\Lambda (0)$ is infinite, $u(0)$ tends to zero, and no information can be obtained.  This rate depends naturally on coupling between the qubits and the open waveguide, as well as on strength of the modulation.

The interplay between phase and amplitude information in Eq. (\ref{8}) is shown in Fig. \ref{fig:Amp_vs_phase_tomo}, where the difference $\left| {\beta _1 (t)} \right|^2  - \left| {\beta _3 (t)} \right|^2$ is taken in the limit $ e^{ - \Lambda(t)}  \to 1$ .
We suppress the first term by choosing $u(t)=\pi/2$ and from Fig. \ref{fig:Amp_vs_phase_tomo} one sees that for any initial phase difference between qubit states $\left| 1 \right\rangle$ and $\left| 3 \right\rangle$ there is a unique value of the population differences.  We also note that in limit when $\left| {\beta _3 (t)} \right|^2=0,1$ the measurable value equals $0$, which becomes clear from Eq. (\ref{rho}) where off-diagonal elements vanish and the phases are totally uncertain.

\begin{figure}[h]
    \includegraphics[width=8 cm]{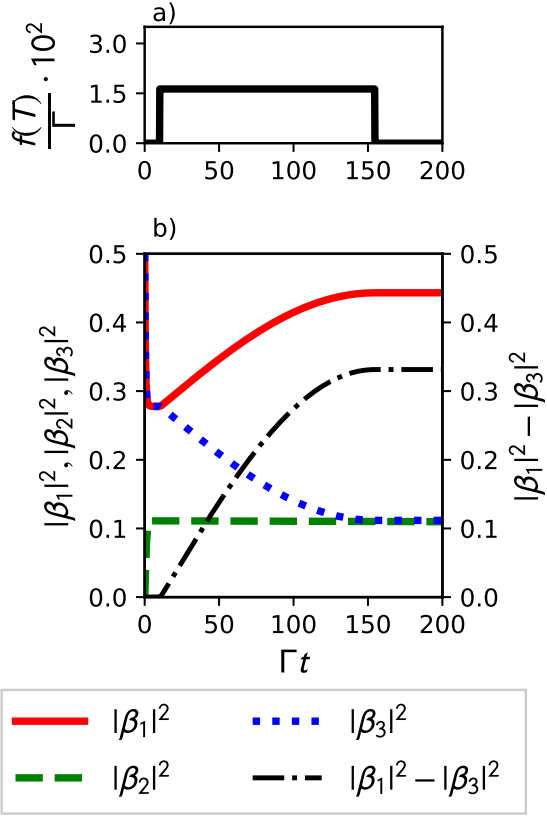}
    \caption{Evolution of probabilities under modulation of the frequency of the central qubit, with parameters  $kd=2\pi$, $u(t)=\pi/2$, $|\beta_1(0)|=|\beta_3(0)|=1/\sqrt{2}, |\beta_2(0)|=0, \varphi_3-\varphi_1=0.4\pi$.
    (a) The modulation is realized as a pulse, starting at $t_0 \cdot \Gamma =10$ and ending at $t_{\rm end} \cdot \Gamma =151$. (b) Probabilities $|\beta_1(t)|^2$ (solid red line), $|\beta_3(t)|^2$ (dotted blue line), $|\beta_2(t)|^2$ (dashed green line) and population difference $|\beta_1(t)|^2-|\beta_3(t)|^2$ (dashdotted black line).}\label{FIG4A}
\end{figure}

As a demonstration of our method we verified the validity of Eq. (\ref{8}) by  numerical simulation for initially equal probability amplitudes $|\beta_1(0)|=|\beta_3(0)|=1/\sqrt{2}$, $\beta_2(0)=0$, and $\varphi_3-\varphi_1=0.4\pi$. In this case, the only non zero term in the right hand side of Eq. (\ref{8}) is proportional to $\sin{u(t)}$. For a $\pi/2$ modulation ($u(t)=\pi/2$) the population difference at the end of the pulse is proportional to $\sin{(\varphi_1-\varphi_3)}$ as it follows from (\ref{11}). This behavior is shown in Fig. \ref{FIG4A}.

\begin{figure}
    \includegraphics[width=8 cm]{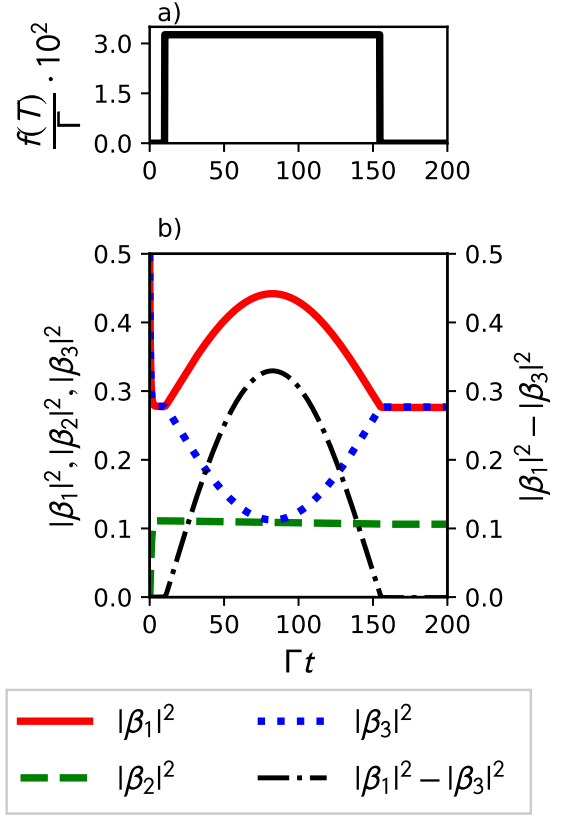}
    \caption{Evolution of probabilities under modulation of the frequency of the central qubit, with parameters $kd=2\pi$, $u(t)=\pi$, $|\beta_1(0)|=|\beta_3(0)|=1/\sqrt{2}, |\beta_2(0)|=0, \varphi_3-\varphi_1=0.4\pi$.
    (a) The modulation is realized as a pulse, starting at $t_0 \cdot \Gamma =10$ and ending at $t_{\rm end} \cdot \Gamma =151$. (b) Probabilities $|\beta_1(t)|^2$ (solid red line), $|\beta_3(t)|^2$ (dotted blue line), $|\beta_2(t)|^2$ (dashed green line) and population difference $|\beta_1(t)|^2-|\beta_3(t)|^2$ (dashdotted black line).}\label{FIG4B}
\end{figure}

Alternatively, for a $\pi$ modulation pulse ($u(t)=\pi$) the population difference $|\beta_1(t)|^2-|\beta_3(t)|^2$ after the end of the pulse becomes equal to zero which is shown in Fig. \ref{FIG4B}.

Also, it is worth mentioning that after the modulation pulse $\left| {\beta _1 (t)} \right|^2  + \left| {\beta _3 (t)} \right|^2 \ne 1$, because the central qubit becomes partially excited. Nevertheless, we are interested only in a combination of measured populations.
In summary to this section, we may conclude that modulating the frequency of the second qubit allows us to obtain
the unambiguous information about the phase difference $\varphi_1-\varphi_3$
via the measurement of the evolution of the probability amplitudes
$|\beta_1(t)|^2$, $|\beta_3(t)|^2$.

 To emulate the reconstruction procedure  we take  the state with an unknown phase difference ${\varphi _1} - {\varphi _3}$ in range $-\pi$ to $\pi$  and with unknown populations $\beta_{1,3}$.
Then, we simulate the dynamics after the $u=\pi$ pulse and get the difference of populations $d(t_{\pi})=\left|{\beta _{1}} \right|^2-\left|{\beta _{3}} \right|^2$. The estimation of the populations from Eq. (\ref{8a}) is:

\begin{equation}
\label{Amplitude_estimation}
    \begin{gathered}
     \left|\beta _1^{est}\left( 0 \right) \right|= \sqrt {\frac{1}{2}\left[ {1 - 3d\left( {{t_\pi }} \right)} \right]} ,\\
      \left| \beta _3^{est}\left( 0 \right)\right| = \sqrt {\frac{1}{2}\left[ {1 + 3d\left( {{t_\pi }} \right)} \right]}.
    \end{gathered}
\end{equation}

At the next step we simulate the dynamics after a $u=\pi /2$ pulse and take the populations $\left|{\beta _{1}} \right|^2$ and $\left| {\beta _{3}} \right|^2$ after the pulse. Then, following equations Eqs. (\ref{11}) and (\ref{8_b}), where $S$ and $d(t_{\pi/2})$ are in fact the measured values, we find the $\sin$ and $\cos$ values of estimated phase $\varphi_{\rm est}$:
\begin{equation}
\label{Sincos}
 \begin{gathered}
\begin{array}{l}
\sin \left( {{\varphi _{est}}} \right) = \frac{3}{2}\frac{{d\left( {{t_{{\pi  \mathord{\left/
 {\vphantom {\pi  2}} \right.
 \kern-\nulldelimiterspace} 2}}}} \right)}}{{\left| {\beta _1^{est}\left( 0 \right)} \right|\left| {\beta _3^{est}\left( 0 \right)} \right|}},\\
\\
\cos \left( {{\varphi _{est}}} \right) =  - \left( {S - \frac{5}{9}} \right)\frac{9}{{8\left| {\beta _1^{est}\left( 0 \right)} \right|\left| {\beta _3^{est}\left( 0 \right)} \right|}}
\end{array}
\end{gathered}
\end{equation}
which allows to explicitly get $\varphi_{\rm est}$  through arctangent. These two steps are enough to reconstruct the state in the form Eq. (\ref{rho}).

\section{Conclusion}

In this paper we have considered three non interacting qubits
embedded in an open waveguide. For this system we have described
experimentally accessible method for the reconstruction within a
single-excitation subspace of arbitrary two-qubit state. The
method is based on the modulation of the frequency of a central
ancillary qubit which allows us to determine the elements of
reduced density matrix for two edge qubits.

In contrast to a common quantum tomography reconstruction where
the gate pulses are applied to the measured qubits, in our method
the measurement pulse is applied to the ancillary qubit. Until the
projective measurements two edge qubits do not undergo any
external influence.

\textbf{Acknowledgements}
Ya. S. G. and A. A. Sh. acknowledge the support from the Ministry of Science
and Higher Education of Russian Federation under grant FSUN-2020-0004.


\appendix
\section{Derivation of the equations for the probability amplitudes
$\beta_n(t)$}

The equations for probability amplitudes $\beta_n(t)$ of the qubits and that of the photon $\gamma_k(t)$  from
Eq. (\ref{eq:math:a4}) can be found from the time-dependent Schrodinger
equation $ id\left| \Psi  \right\rangle /dt = H\left| \Psi
\right\rangle $ . For the amplitudes  we obtain:

\begin{equation}\label{bet1}
\frac{{d\beta _1 }}
{{dt}} =  - i\sum\limits_k {} g_k^{} e^{ - ikd} \gamma _k (t)e^{ - i(\omega _k  - \Omega )t} ,
\end{equation}
\begin{equation}\label{bet2}
\frac{{d\beta _2 }}
{{dt}} =  - if(t)\beta _2 (t) - i\sum\limits_k {} g_k^{} \gamma _k (t)e^{ - i(\omega _k  - \Omega )t} ,
\end{equation}
\begin{equation}\label{bet3}
\frac{{d\beta _3 }}
{{dt}} =  - i\sum\limits_k {} g_k^2 e^{ikd} \gamma _k (t)e^{ - i(\omega _k  - \Omega )t} ,
\end{equation}
\begin{equation}\label{gam1}
\frac{{d\gamma _k (t)}}
{{dt}} =  - ig_k F_k (t)e^{i(\omega _k  - \Omega )t} ,
\end{equation}
where
\begin{equation}\label{A3}
F_k (t) = \beta _1 (t)e^{ikd}  + \beta _2 (t) + \beta _3 (t)e^{ -
ikd} .
\end{equation}
From (\ref{gam1}) we obtain:

\begin{equation}\label{A2}
\gamma _k (t) =  - ig_k \int\limits_0^t {} F_k (t')e^{  i(\omega
_k  - \Omega )t'} dt' .
\end{equation}
The expression (\ref{A2}) allows us to remove the photon amplitude $\gamma_k(t)$
from the equations for the qubits' amplitudes (\ref{bet1}), (\ref{bet2}), and (\ref{bet3}). The result is as follows:
\begin{equation}\label{A1}
\begin{gathered}
  \frac{{d\beta _1 }}
{{dt}} =  - \sum\limits_k {} g_k^2 e^{ - ikd} \int\limits_0^t {} F_k (t')e^{ - i(\omega _k  - \Omega )(t - t')} dt' ,\hfill \\
  \frac{{d\beta _2 }}
{{dt}} =  - if(t)\beta _2 (t) - \sum\limits_k {} g_k^2 \int\limits_0^t {} F_k (t')e^{ - i(\omega _k  - \Omega )(t - t')} dt' ,\hfill \\
  \frac{{d\beta _3 }}
{{dt}} =  - \sum\limits_k {} g_k^2 e^{ikd} \int\limits_0^t {} F_k (t')e^{ - i(\omega _k  - \Omega )(t - t')} dt' ,\hfill \\
\end{gathered}
\end{equation}

In accordance with Wigner-Weiskopff approximation we take the
quantity $F_k(t)$ out the integrands,
\begin{equation}\label{A4}
\begin{gathered}
  \frac{{d\beta _1 }}
{{dt}} =  - \sum\limits_k {} g_k^2 e^{ - ikd} F_k (t)I_k (\Omega ,t) , \hfill \\
  \frac{{d\beta _2 }}
{{dt}} =  - if(t)\beta _2 (t) - \sum\limits_k {} g_k^2 F_k (t)I_k (\Omega ,t) ,\hfill \\
  \frac{{d\beta _3 }}
{{dt}} =  - \sum\limits_k {} g_k^2 e^{ikd} F_k (t)I_k (\Omega ,t) ,\hfill \\
\end{gathered}
\end{equation}

where

\begin{equation}\label{A5}
\begin{gathered}
  I_k (\Omega ,t) = \int\limits_0^t {e^{ - i(\omega _k  - \Omega )(t - t')} dt'}  = \int\limits_0^t {e^{ - i(\omega _k  - \Omega )\tau } d\tau }  \hfill \\
   \approx \int\limits_0^\infty  {e^{ - i(\omega _k  - \Omega )\tau } d\tau }  = \pi \delta (\omega _k  - \Omega ) - iP.v.\left( {\frac{1}
{{\omega _k  - \Omega }}} \right) \hfill \\
\end{gathered}
\end{equation}

We assume $g_{-k}=g_k, I_{-k}(\Omega,t)=I_{k}(\Omega,t)$ and leave
the summation in (\ref{A4}) over positive valued of $k$ (positive
frequencies).

\begin{equation}\label{A6}
\begin{gathered}
  \frac{{d\beta _1 }}
{{dt}} =  - \sum\limits_{k > 0} {} g_k^2 \left( {e^{ - ikd} F_k (t) + e^{ikd} F_{ - k} (t} \right)I_k (\Omega ,t), \hfill \\
  \frac{{d\beta _2 }}
{{dt}} =  - if(t)\beta _2 (t) - \sum\limits_{k > 0} {} g_k^2 \left( {F_k (t) + F_{ - k} (t} \right)I_k (\Omega ,t), \hfill \\
  \frac{{d\beta _3 }}
{{dt}} =  - \sum\limits_{k > 0} {} g_k^2 \left( {e^{ikd} F_k (t) + e^{ - ikd} F_{ - k} (t} \right)I_k (\Omega ,t). \hfill \\
\end{gathered}
\end{equation}

Inserting the explicit form of $F_k(t)$ (\ref{A3}) in (\ref{A6}) results in
the following expressions:

\begin{equation}\label{A7}
\begin{gathered}
  \frac{{d\beta _1 }}
{{dt}} =  - 2\sum\limits_{k > 0} {} g_k^2 \left( {\beta _1  + \beta _2 \cos kd + \beta _3 \cos 2kd} \right)I_k (\Omega ,t) ,\hfill \\
  \frac{{d\beta _2 }}
{{dt}} =  - if(t)\beta _2 (t) \hfill \\
   - 2\sum\limits_{k > 0} {} g_k^2 \left( {\beta _1 \cos kd + \beta _2  + \beta _3 \cos kd} \right)I_k (\Omega ,t) ,\hfill \\
  \frac{{d\beta _3 }}
{{dt}} =  - 2\sum\limits_{k > 0} {} g_k^2 \left( {\beta _1 \cos 2kd + \beta _2 \cos kd + \beta _3 } \right)I_k (\Omega ,t) .\hfill \\
\end{gathered}
\end{equation}

The next step is to relate the coupling constants $g_k$ to the
qubit decay rate of spontaneous emission into the waveguide mode. In
accordance with Fermi golden rule we define the qubit decay rates
by the following expressions:

\begin{equation}\label{A8}
\Gamma  = 2\pi \sum\limits_k {g_k^2 \delta (\omega _k  - \Omega )}
\end{equation}

For the 1D case, a summation over $k$ is replaced by an integration
over $\omega$  in accordance with the prescription

\begin{equation}\label{A9}
2\sum\limits_{k > 0} {}  \Rightarrow \frac{L} {{2\pi
}}2\int\limits_0^\infty  {d\left| k \right|}  = \frac{L} {{\pi
\upsilon _g }}\int\limits_0^\infty  {d\omega _k },
\end{equation}

where $L$ is a length of the waveguide, and we assumed a linear
dispersion law,  $\omega_k=v_g|k|$. The application of (\ref{A9})
to (\ref{A8}), allows to derive a  relation between the coupling constant
$g_k$ and the decay rate $\Gamma$,

\begin{equation}\label{A10}
g_k^{}  = \left( {\frac{{v_g \Gamma }} {{2L}}} \right)^{1/2}.
\end{equation}

Now we can calculate the different terms in (\ref{A7}).
\begin{equation}\label{A11}
\begin{gathered}
  \sum\limits_k {g_k^2 I_k (\Omega ,t)}  = \sum\limits_k {g_k^2 } \left( {\pi \delta (\omega _k  - \Omega ) - iP.v.\left( {\frac{1}
{{\omega _k  - \Omega }}} \right)} \right) \hfill \\
   = \frac{\Gamma }
{2} - iP.v.\left( {\frac{{g_k^2 }} {{\omega _k  - \Omega }}}
\right) \approx \frac{\Gamma }
{2}, \hfill \\
\end{gathered}
\end{equation}

\begin{equation}\label{A12}
\begin{gathered}
  2\sum\limits_{k > 0} {g_k^2\cos (kd)} I_k (\Omega ,t) \hfill \\
   = \frac{L}
{{\upsilon _g }}\int\limits_0^\infty  {g_k^2 \cos (kd)\delta
(\omega _k  - \Omega )d\omega _k }  - 2i\sum\limits_{k > 0} {}
P.v.\left( {\frac{{g_k^2 \cos (kd)}}
{{\omega _k  - \Omega _{} }}} \right) \hfill \\
   = \frac{L}
{{\upsilon _g }}g_\Omega ^2 \cos (k_\Omega  d) - i\frac{L} {{v_g
\pi }}g_\Omega ^2 P.v.\int\limits_0^\infty  {} \frac{{\cos \left(
{\frac{\omega } {{v_g }}d} \right)}}
{{\omega  - \Omega }} \hfill \\
\end{gathered}
\end{equation}

For the principal value integral in (\ref{A12}) we obtain:
\begin{equation}\label{A13}
P.v.\int\limits_0^\infty  {d\omega} \frac{{\cos \left(
{\frac{\omega } {{v_g }}d} \right)}} {{\omega  - \Omega }} =-\pi
\sin \left( {k_\Omega  d} \right)
\end{equation}
where $k_\Omega=\Omega/v_g$.

The expression (\ref{A13}) is exact if  counter-rotating terms in
the qubit-field interaction are taken into account (Suppl. in
\cite{Gonz2013}). Nevertheless, within a rotating wave
approximation the Eq. \ref{A13} provides a good accuracy for
$d>\lambda/4$ \cite{Green2021}.

\begin{equation}\label{A14}
2\sum\limits_{k > 0} {g_k^2 \cos (kd)} I_k (\Omega ,t) = \frac{L}
{{\upsilon _g }}g_\Omega ^2 e^{ik_\Omega  d}  = \frac{\Gamma }
{2}e^{ik_\Omega  d}.
\end{equation}

Similar calculations also give for the sum in (\ref{A7}):

\begin{equation}\label{A15}
2\sum\limits_{k > 0} {g_k^2 \cos (2kd)} I_k (\Omega ,t) =
\frac{\Gamma } {2}e^{2ikd} .
\end{equation}

In (\ref{A11}) the decay rate $\Gamma$ is defined by (\ref{A8}).
The principal value in (\ref{A11}) gives rise to the shift of the
qubit frequency. Therefore, we incorporate it in the renormalized
qubit frequency and will not write it explicitly any more.

Collecting together (\ref{A11}), (\ref{A14}), and (\ref{A15}) we
write the final form of the equations (\ref{A7}):

\begin{equation}\label{A16}
\begin{gathered}
  \frac{{d\beta _1 }}
{{dt}} =  - \frac{\Gamma }
{2}\left( {\beta _1  + \beta _2 e^{ikd}  + \beta _3 e^{i2kd} } \right) , \hfill \\
  \frac{{d\beta _2 }}
{{dt}} =  - if(t)\beta _2 (t) - \frac{\Gamma }
{2}\left( {\beta _1 e^{ikd}  + \beta _2  + \beta _3 e^{ikd} } \right) ,\hfill \\
  \frac{{d\beta _3 }}
{{dt}} =  - \frac{\Gamma }
{2}\left( {\beta _1 e^{i2kd}  + \beta _2 e^{ikd}  + \beta _3 } \right) , \hfill \\
\end{gathered}
\end{equation}

\section{Derivation of equation (\ref{8})}
Equations (\ref{5}) can be written in the matrix form:
\begin{equation}\label{B1}
\frac{d\widehat{\beta}}{dt}=A(t)\widehat{\beta}(t),
\end{equation}
where
\begin{equation}\label{B2}
{\widehat{\beta}(t){  = }}\left( {\begin{array}{*{20}c}
   {\beta _1(t) }  \\
   {\beta _2 }(t)  \\
   {\beta _3(t) }  \\
\end{array} } \right),
\end{equation}

\begin{equation}\label{B3}
{\text{A(t) = }} - \frac{\Gamma } {2}\left( {\begin{array}{*{20}c}
   1 & {e^{ikd} } & {e^{2ikd} }  \\
   {e^{ikd} } & {1 + if(t)\frac{2}
{\Gamma }} & {e^{ikd} }  \\
   {e^{2ikd} } & {e^{ikd} } & 1  \\
\end{array} } \right) .
\end{equation}

It is easy to verify that the matrices $A(t)$ do not commute at
different times $[A(t_1),A(t_2)]\neq 0$. In this case the solution
of (\ref{B1}) can be obtained in the form:
\begin{equation}\label{B4}
\widehat{\beta} (t) = e^{M(t)} \widehat{\beta} (0)
\end{equation}

where the Magnus operator $M(t)$ can be written as infinite series
expansion \cite{Blanes2009}:
\begin{equation}\label{B5}
M(t) = \sum\limits_{n = 1}^\infty  {M_n } (t).
\end{equation}

The first two terms in (\ref{B5}) are as follows:
\begin{equation}\label{B6}
\begin{gathered}
  M_1 (t) = \int\limits_0^t {dt_1 A(t_1 )} , \hfill \\
  M_2 (t) = \frac{1}
{2}\int\limits_0^t {dt_2 \int\limits_0^{t_2 } {dt_1 } \left[ {A(t_1 )A(t_2 )} \right]} . \hfill \\
\end{gathered}
\end{equation}

According to Silvester's matrix theorem (named after J. J.
Sylvester) any analytic function $z(M)$ of a quadratic $n\times n$
matrix $M$ can be expressed as a polynomial in $M$, in terms of
the eigenvalues and eigenvectors of $M$ \cite{Horn1991}. Specifically, the theorem states that

\begin{equation}\label{B7}
z(M) = \sum\limits_{i = 1}^n {z(\lambda _i )B_i },
\end{equation}

where $\lambda_i$ are the characteristic roots of the equation

\begin{equation}\label{B8}
\det \left| {M - \lambda I} \right| = 0 ,
\end{equation}

and
\begin{equation}\label{B9}
B_i  = \prod\limits_{j = 1,i \ne j}^n {\frac{{M - \lambda _j I}}
{{\lambda _i  - \lambda _j }}},
\end{equation}
where $I$ is the identity matrix.

The Silvester's formula (\ref{B7}) holds for any quadratic
diagonalizable matrix all roots of which are different.

In the sum of (\ref{B5}) we neglect all terms except for the first
one, $M(t)=M_1(t)$:
\begin{equation}\label{B10}
M_1 (t) = \int\limits_0^t {dt_1 A(t_1 )} {\text{ = }} -
\frac{{\Gamma t}} {2}\left( {\begin{array}{*{20}c}
   1 & {e^{ikd} } & {e^{2ikd} }  \\
   {e^{ikd} } & {\Omega (t)} & {e^{ikd} }  \\
   {e^{2ikd} } & {e^{ikd} } & 1  \\
 \end{array} } \right),
\end{equation}

where
\begin{equation}\label{B11}
\Omega (t) = 1 + i\frac{2} {{\Gamma t}}\int\limits_0^t {f(t_1
)dt_1 }  = 1 - \frac{2} {{\Gamma t}}F(t) ,
\end{equation}

\begin{equation}\label{B12}
F(t) =  - i\int\limits_0^t {f(t_1 )dt_1 }.
\end{equation}

Next, we find the characteristic roots $\lambda_i(t)$  of the
matrix $M_1(t)$, which are the roots of the equation
\begin{equation}\label{B13}
\det \left| {M_1(t) - \lambda (t)I} \right| = 0 .
\end{equation}
The equation (\ref{B13}) is a cubic equation
\begin{equation}\label{B14}
\left( {1 - \lambda } \right)^2 \left( {\Omega  - \lambda }
\right) + 2e^{4ikd}  - e^{4ikd} \left( {\Omega  - \lambda }
\right) - 2e^{2ikd} \left( {1 - \lambda } \right) = 0 ,
\end{equation}
with the following three roots
\begin{equation}\label{r1}
\begin{gathered}
  \lambda _{1,2} (t) =  - \frac{{\Gamma t}}
{2}\left( {1 + \frac{1} {2}e^{2ikd} } \right) + \frac{1}
{2}F(t) \hfill \\
   \pm \frac{1}
{4}e^{ikd} \sqrt {\left( {8 + e^{2ikd} } \right)\Gamma ^2 t^2  + 4F^2 (t)e^{ - 2ikd}  + 4F(t)\Gamma t}  ,\hfill \\
\end{gathered}
\end{equation}
\begin{equation}\label{r3}
\lambda _3 (t) = \frac{{\Gamma t}} {2}\left( {e^{2ikd}  - 1}
\right).
\end{equation}

In the equation (\ref{r1}) the roots $\lambda_1$, $\lambda_2$
correspond to $+$, $-$ sign, respectively.

 The application of \ref{B7} to $z(M)=e^M$ gives rise to the
following equation:
\begin{equation}\label{B18}
e^{M_1(t)}  = B_1 e^{\lambda _1 }  + B_2 e^{\lambda _2 }  + B_3
e^{\lambda _3 },
\end{equation}

where
\begin{equation}\label{B19}
\begin{gathered}
  B_1  = \left( {\frac{{M_1 - \lambda _2 I}}
{{\lambda _1  - \lambda _2 }}} \right)\left( {\frac{{M_1 - \lambda
_3 I}}
{{\lambda _1  - \lambda _3 }}} \right) \hfill \\
 \quad  = \frac{{M_{1}^2  - (\lambda _2  + \lambda _3 )M_1 + \lambda _2 \lambda _3 I}}
{{(\lambda _1  - \lambda _2 )(\lambda _1  - \lambda _3 )}}, \hfill \\
\end{gathered}
\end{equation}

\begin{equation}\label{B20}
\begin{gathered}
  B_2  = \left( {\frac{{M_1 - \lambda _1 I}}
{{\lambda _2  - \lambda _1 }}} \right)\left( {\frac{{M_1 - \lambda
_3 I}}
{{\lambda _2  - \lambda _3 }}} \right) \hfill \\
  \quad  = \frac{{M_{1}^2  - (\lambda _1  + \lambda _3 )M_1 + \lambda _1 \lambda _3 I}}
{{(\lambda _2  - \lambda _1 )(\lambda _2  - \lambda _3 )}} ,\hfill \\
\end{gathered}
\end{equation}

\begin{equation}\label{B21}
\begin{gathered}
  B_3  = \left( {\frac{{M_1 - \lambda _1 I}}
{{\lambda _3  - \lambda _1 }}} \right)\left( {\frac{{M_1 - \lambda
_2 I}}
{{\lambda _3  - \lambda _2 }}} \right) \hfill \\
  \quad  = \frac{{M_{1}^2  - (\lambda _1  + \lambda _2 )M_1 + \lambda _2 \lambda _1 I}}
{{(\lambda _3  - \lambda _1 )(\lambda _3  - \lambda _2 )}} .\hfill \\
\end{gathered}
\end{equation}
The equations (\ref{B18})-(\ref{B21}) are valid for any value of
$kd$.

Below we assume $kd=\pi n$, where $n$ is a positive integer. For
this  case $\lambda_3=0$ and we obtain
\begin{equation}\label{B22}
e^{M_1(t)}  = B_1 e^{\lambda _1 }  + B_2 e^{\lambda _2 }  + B_3 ,
\end{equation}

\begin{equation}\label{B23}
\begin{gathered}
  B_1  = \frac{{M_{1}^2  - \lambda _2 M_1}}
{{(\lambda _1  - \lambda _2 )\lambda _1 }} ,\hfill \\
  B_2  = \frac{{M_{1}^2  - \lambda _1 M_1}}
{{(\lambda _2  - \lambda _1 )\lambda _2 }} ,\hfill \\
  B_3  = \frac{{M_{1}^2  - (\lambda _1  + \lambda _2 )M_1}}
{{\lambda _1 \lambda _2 }} + I , \hfill \\
\end{gathered}
\end{equation}
where
\begin{equation}\label{B24}
\begin{gathered}
  \lambda _{1,2} (t) =  - \frac{{3\Gamma t}}
{4} + \frac{1}
{2}F(t) \hfill \\
  \quad \quad  \pm ( - 1)^n \frac{1}
{4}\sqrt {9\Gamma ^2 t^2  + 4F^2 (t) + 4F(t)\Gamma t} . \hfill \\
\end{gathered}
\end{equation}

Below we perform the calculations for $kd=2\pi$. Using the
equation (\ref{B4})  and the explicit expression (\ref{B10}) for
the matrix $M_1(t)$ we obtain from (\ref{B22}), (\ref{B23}),
(\ref{B24}) the expressions for qubits amplitudes $\beta_1(t)$,
$\beta_3(t)$.
\begin{widetext}
\begin{equation}\label{B25}
\beta _1 (t,2\pi ) = \frac{{\Gamma t}} {R}\left( {\beta _1 (0) +
\beta _3 (0)} \right)\left( {\frac{{\left( {\frac{{3\Gamma t}} {2}
+ F - \frac{R} {2}} \right)}} {{\left( { - \frac{{3\Gamma t}} {2}
+ F + \frac{1} {2}R} \right)}}e^{\lambda _1 }  - \frac{{\left(
{\frac{3} {2}\Gamma t + F + \frac{R} {2}} \right)}} {{\left( { -
\frac{{3\Gamma t}} {2} + F - \frac{1} {2}R} \right)}}e^{\lambda _2
} } \right) + \frac{1} {2}\left( {\beta _1 (0) - \beta _3 (0)}
\right),
\end{equation}

\begin{equation}\label{B26}
\beta _3 (t,2\pi ) = \frac{{\Gamma t}} {R}\left( {\beta _1 (0) +
\beta _3 (0)} \right)\left( {\frac{{\left( {\frac{{3\Gamma t}} {2}
+ F - \frac{R} {2}} \right)}} {{\left( { - \frac{{3\Gamma t}} {2}
+ F + \frac{1} {2}R} \right)}}e^{\lambda _1 }  - \frac{{\left(
{\frac{3} {2}\Gamma t + F + \frac{R} {2}} \right)}} {{\left( { -
\frac{{3\Gamma t}} {2} + F - \frac{1} {2}R} \right)}}e^{\lambda _2
} } \right) - \frac{1} {2}\left( {\beta _1 (0) - \beta _3 (0)}
\right),
\end{equation}

\end{widetext}

where
\begin{equation}\label{B27}
R = \sqrt {9\Gamma ^2 t^2  + 4F^2 (t) + 4F(t)\Gamma t} .
\end{equation}

Now we analyze the quantities $R$, $\lambda_1$, $\lambda_2$ for
$\Gamma t\gg |F|$. We obtain
\begin{equation}\label{B28}
R \approx 3\Gamma t + \frac{2} {3}F + \frac{{16}} {{27}}\frac{{F^2
}} {{\Gamma t}},
\end{equation}

\begin{equation}\label{B29}
\lambda _1 (t) =  - \frac{{3\Gamma t}} {4} + \frac{1} {2}F(t) +
\frac{1} {4}R \approx \frac{2} {3}F + \frac{4} {{27}}\frac{{F^2 }}
{{\Gamma t}},
\end{equation}

\begin{equation}\label{B30}
\lambda _2 (t) =  - \frac{{3\Gamma t}} {4} + \frac{1} {2}F(t) -
\frac{1} {4}R \approx  - \frac{3} {2}\Gamma t + \frac{1} {3}F -
\frac{4} {{27}}\frac{{F^2 }} {{\Gamma t}}.
\end{equation}

Therefore, in (\ref{B25}), (\ref{B26}) we neglect the decaying
exponent $ e^{\lambda _2 }  \approx e^{ - \frac{3} {2}\Gamma t} $.
The quantity $e^{\lambda_1}$ we write in the following form:
\begin{equation}\label{B31}
e^{\lambda _1 }  \approx \exp \left( {\frac{2} {3}F + \frac{4}
{{27}}\frac{{F^2 }} {{\Gamma t}}} \right) \equiv e^{ - iu(t)} e^{
- \Lambda (t)},
\end{equation}

where
\begin{equation}\label{B31a}
u(t) = \frac{2} {3}\int\limits_0^t {f(\tau )d\tau },
\end{equation}

\begin{equation}\label{B32}
\Lambda (t) = \frac{4} {{27\Gamma t}}\left( {\int\limits_0^t
{f(\tau )d\tau } } \right)^2 .
\end{equation}

For this approximation the equations (\ref{B25}), (\ref{B26}) take
the form:

\begin{equation}\label{B33}
\begin{gathered}
  \beta _1 (t,2\pi ) = \frac{1}
{6}\left( {\beta _1 (0) + \beta _3 (0)} \right)e^{ - iu(t)} e^{ - \Lambda (t)}  \hfill \\
   + \frac{1}
{2}\left( {\beta _1 (0) - \beta _3 (0)} \right) ,\hfill \\
\end{gathered}
\end{equation}

\begin{equation}\label{B34}
\begin{gathered}
  \beta _3 (t,2\pi ) = \frac{1}
{6}\left( {\beta _1 (0) + \beta _3 (0)} \right)e^{ - iu(t)} e^{ - \Lambda (t)}  \hfill \\
   - \frac{1}
{2}\left( {\beta _1 (0) - \beta _3 (0)} \right) ,\hfill \\
\end{gathered}
\end{equation}

\begin{equation}\label{B34a}
\beta _2 (t) =  - \frac{{\left( {\beta _1 (0) + \beta _3 (0)} \right)}}
{3}e^{ - iu(t)} e^{ - \Lambda (t)} .
\end{equation}

From (\ref{B33}), (\ref{B34}) we finally obtain for $kd=2\pi$
\begin{equation}\label{B35}
\begin{gathered}
  \left| {\beta _1 (t)} \right|^2  - \left| {\beta _3 (t)} \right|^2  = \frac{1}
{3}\left( {\left| {\beta _1 (0)} \right|^2  - \left| {\beta _3 (0)} \right|^2 } \right)e^{ - \Lambda (t)} \cos u(t) \hfill \\
   - \frac{1}
{3}i\left( {\beta _1^ *  (0)\beta _3 (0) - \beta _1 (0)\beta _3^ *  } \right)e^{ - \Lambda (t)} \sin u(t) , \hfill \\
\end{gathered}
\end{equation}

\begin{equation}\label{B36}
\begin{gathered}
  \left| {\beta _1 (t)} \right|^2  + \left| {\beta _3 (t)} \right|^2  = \frac{1}
{{18}}\left( {e^{ - \Lambda (t)}  + 9} \right) \hfill \\
   + \frac{1}
{9}\left( {\beta _1^ *  (0)\beta _3 (0) + \beta _1 (0)\beta _3^ *  } \right)\left( {e^{ - \Lambda (t)}  - 9} \right) ,\hfill \\
\end{gathered}
\end{equation}

\begin{equation}\label{B38}
\left| {\beta _2 (t)} \right|^2  = \frac{1}
{9}e^{ - 2\Lambda (t)} \left( {1 + 2\left| {\beta _1 (0} \right|\left| {\beta _3 (0)} \right|\cos \left( {\varphi _1  - \varphi _3 } \right)} .\right)
\end{equation}
which are the equations (\ref{8}), (\ref{8a}), (\ref{8b}) from the main text.

\end{document}